\documentclass[twoside,twocolumn]{article}
\usepackage[textwidth=7.5in,textheight=8.7in]{geometry}

\usepackage[sc]{mathpazo} 
\usepackage[T1]{fontenc} 
\usepackage{microtype} 
\usepackage{amsmath}
\usepackage{upgreek}
\usepackage{environ}
\usepackage{float}
\usepackage{todonotes}
\NewEnviron{myequation}{%
\begin{equation}
\scalebox{1.5}{$\BODY$}
\end{equation}

}

\usepackage{amssymb}
\usepackage{array,siunitx}

\usepackage{algorithm}
\usepackage{algpseudocode}
\algnewcommand\And{\textbf{and} }
\algnewcommand{\Or}{\textbf{or}}
\newcommand{\vars}{\texttt}
\usepackage{subfig}
\DeclareMathSizes{12}{30}{16}{12}
\usepackage{textcomp}
\usepackage{gensymb}
\usepackage[LGRgreek]{mathastext}
\algnewcommand\algorithmicforeach{\textbf{for each}}
\algdef{S}[FOR]{ForEach}[1]{\algorithmicforeach\ #1\ \algorithmicdo}



\RequirePackage{tikz}
\RequirePackage{zref-abspage}
\RequirePackage{zref-user}
\RequirePackage{tikz}
\RequirePackage{atbegshi}
\usetikzlibrary{calc}
\RequirePackage{tikzpagenodes}
\RequirePackage{etoolbox}

\makeatletter
\newcommand*\ALG@lastblockb{b}
\newcommand*\ALG@lastblocke{e}
\apptocmd{\ALG@beginblock}{%
    \ifx\ALG@lastblock\ALG@lastblockb
        \ifnum\theALG@nested>1\relax\expandafter\@firstoftwo\else\expandafter\@secondoftwo\fi{\ALG@tikzborder}{}%
    \fi
    \let\ALG@lastblock\ALG@lastblockb%
}{}{\errmessage{failed to patch}}

\pretocmd{\ALG@endblock}{%
    \ifx\ALG@lastblock\ALG@lastblocke
        \addtocounter{ALG@nested}{1}%
        \addtolength\ALG@tlm{\csname ALG@ind@\theALG@nested\endcsname}%
        \ifnum\theALG@nested>1\relax\expandafter\@firstoftwo\else\expandafter\@secondoftwo\fi{\endALG@tikzborder}{}%
        \addtolength\ALG@tlm{-\csname ALG@ind@\theALG@nested\endcsname}%
        \addtocounter{ALG@nested}{-1}%
    \fi
    \let\ALG@lastblock\ALG@lastblocke%
}{}{\errmessage{failed to patch}}

\tikzset{ALG@tikzborder/.style={line width=0.5pt,black}}

\newcommand*\currenttextarea{current page text area}

\newcommand*{\updatecurrenttextarea}{%
    \if@twocolumn
        \if@firstcolumn
            \renewcommand*{\currenttextarea}{current page column 1 area}%
        \else
            \renewcommand*{\currenttextarea}{current page column 2 area}%
        \fi
    \else
        \renewcommand*\currenttextarea{current page text area}%
    \fi
}

\newcounter{ALG@tikzborder}
\newcounter{ALG@totaltikzborder}
\newenvironment{ALG@tikzborder}[1][]{%
    \ifx&#1&\else
        \tikzset{ALG@tikzborder/.style={#1}}%
    \fi
    \stepcounter{ALG@totaltikzborder}%
    \expandafter\edef\csname ALG@ind@border@\theALG@nested\endcsname{\theALG@totaltikzborder}%
    \setcounter{ALG@tikzborder}{\csname ALG@ind@border@\theALG@nested\endcsname}%
    \tikz[overlay,remember picture] \coordinate (ALG@tikzborder-\theALG@tikzborder);
    \zlabel{ALG@tikzborder-begin-\theALG@tikzborder}%
    \ifnum\zref@extract{ALG@tikzborder-begin-\theALG@tikzborder}{abspage}=\zref@extract{ALG@tikzborder-end-\theALG@tikzborder}{abspage} \else
        \updatecurrenttextarea
        \ALG@drawvline{[shift={(0pt,0.5\ht\strutbox)}]ALG@tikzborder-\theALG@tikzborder}{\currenttextarea.south east}{\ALG@thistlm}%
        \newcounter{ALG@tikzborderpages\theALG@tikzborder}%
        \setcounter{ALG@tikzborderpages\theALG@tikzborder}{\numexpr-\zref@extract{ALG@tikzborder-begin-\theALG@tikzborder}{abspage}+\zref@extract{ALG@tikzborder-end-\theALG@tikzborder}{abspage}}%
        \ifnum\value{ALG@tikzborderpages\theALG@tikzborder}>1
            \edef\nextcmd{\noexpand\AtBeginShipoutNext{\noexpand\ALG@tikzborderpage{\theALG@tikzborder}{\the\ALG@thistlm}}}
            \nextcmd
        \fi
    \fi
}{%
    \setcounter{ALG@tikzborder}{\csname ALG@ind@border@\theALG@nested\endcsname}%
    \tikz[overlay,remember picture] \coordinate (ALG@tikzborder-end-\theALG@tikzborder);
    \zlabel{ALG@tikzborder-end-\theALG@tikzborder}%
    \updatecurrenttextarea
    \ifnum\zref@extract{ALG@tikzborder-begin-\theALG@tikzborder}{abspage}=\zref@extract{ALG@tikzborder-end-\theALG@tikzborder}{abspage}\relax
    \fi
}

\newcommand*{\ALG@drawvline}[3]{
    \begin{tikzpicture}[overlay,remember picture]
        \draw [ALG@tikzborder]
            let \p0 = (\currenttextarea.north west), \p1=(#1), \p2 = (#2)
             in
            (#3+\fboxsep+.5\pgflinewidth+\x0,\y1+\fboxsep+.5\pgflinewidth)
             --
            (#3+\fboxsep+.5\pgflinewidth+\x0,\y2-\fboxsep-.5\pgflinewidth)
        ;
    \end{tikzpicture}%
}

\newcommand{\ALG@tikzborderpage}[2]{
    \updatecurrenttextarea
    \setcounter{ALG@tikzborder}{#1}%
    \ALG@drawvline{\currenttextarea.north west}{\currenttextarea.south east}{#2}%
    \addtocounter{ALG@tikzborderpages\theALG@tikzborder}{-1}%
    \ifnum\value{ALG@tikzborderpages\theALG@tikzborder}>1
        \AtBeginShipoutNext{\ALG@tikzborderpage{#1}{#2}}%
    \fi
}

\def\ALG@tikzbordertext{\the\ALG@tlm}
\makeatother

\RequirePackage{etoolbox}
\makeatletter
\newlength{\ALG@continueindent}
\newcommand*{\ALG@customparshape}{\parshape 2 \leftmargin \linewidth \dimexpr\ALG@tlm+\ALG@continueindent\relax \dimexpr\linewidth+\leftmargin-\ALG@tlm-\ALG@continueindent\relax}
\newcommand*{\ALG@customparshapex}{\parshape 1 \dimexpr\ALG@tlm+\ALG@continueindent\relax \dimexpr\linewidth+\leftmargin-\ALG@tlm-\ALG@continueindent\relax}
\apptocmd{\ALG@beginblock}{\ALG@customparshape\everypar{\ALG@customparshapex}}{}{\errmessage{failed to patch}}
\makeatother

\makeatletter
\renewcommand{\Function}[2]{%
  \csname ALG@cmd@\ALG@L @Function\endcsname{#1}{#2}%
  \def\jayden@currentfunction{#1}%
}
\newcommand{\funclabel}[1]{%
  \@bsphack
  \protected@write\@auxout{}{%
    \string\newlabel{#1}{{\jayden@currentfunction}{\thepage}}%
  }%
  \@esphack
}
\makeatother

\usepackage{mathtools}

\DeclarePairedDelimiter\abs{\lvert}{\rvert}%
\DeclarePairedDelimiter\norm{\lVert}{\rVert}%

\makeatletter
\let\oldabs\abs
\def\abs{\@ifstar{\oldabs}{\oldabs*}}
\let\oldnorm\norm
\def\norm{\@ifstar{\oldnorm}{\oldnorm*}}
\makeatother

\usepackage[bookmarks=false]{hyperref}
    \hypersetup{colorlinks,
      linkcolor=blue,
      citecolor=blue,
      urlcolor=blue}

\usepackage{enumitem} 
\setlist[itemize]{noitemsep} 

\usepackage{abstract} 

\usepackage{titlesec} 
\renewcommand\thesection{\Roman{section}} 
\renewcommand\thesubsection{\roman{subsection}} 
\titleformat{\section}[block]{\large\scshape\centering}{\thesection.}{1em}{} 
\titleformat{\subsection}[block]{\large}{\thesubsection.}{1em}{} 

\usepackage{titling} 

\usepackage{hyperref} 

\providecommand{\keywords}[1]
{
  \small	
  \textbf{\textit{Keywords---}} #1
}

\pretitle{\begin{center}\LARGE} 
\posttitle{\end{center}} 

\title{On Intercultural Transferability and Calibration of Heterogeneous Shared Space Motion Models\thanks{This paper has been published with copyright by Taylor \& Francis Online, accessible at \url{https://www.tandfonline.com/doi/full/10.1080/19427867.2020.1866332}}} 
\author{%
Fatema T. Johora and J\"org P. M\"uller \\[1ex] 
\normalsize Technische Universit\"at Clausthal, Clausthal-Zellerfeld, Germany \\ 
\normalsize \href{mailto:fatema.tuj.johora,joerg.mueller@tu-clausthal.de}{fatema.tuj.johora,joerg.mueller@tu-clausthal.de} 
}
\date{} 

\renewcommand{\maketitlehookd}{
\begin{abstract}
Modelling and simulation of mixed-traffic zones is an important tool for transportation planners to assess safety, efficiency, and human-friendliness of future urban areas. This paper addresses problems of calibration and transferability of existing shared space models when applied to scenarios that differ in terms of cultural aspects, traffic conditions, and spatial layout. In particular, the first contribution of this work is an enhancement of the Game-Theoretic Social Force Model (GSFM) by a generic methodology for largely automated model calibration; we illustrate the use of the calibration method for a shared space environment in Germany. The second contribution is an investigation into transferability of shared space models. We define criteria for model transferability and present a case study, in which we analyse and evaluate transferability of the model we constructed based on the ``German dataset'' to a different shared space environment from China.  Our results indicate that although -- as to be expected -- the model faces difficulties to replicate the movement behaviours of road users from a new environment, by adding social norms (derived through analysis) of that environment to our model, satisfactory improvement of model accuracy can be obtained with limited effort.
\end{abstract}
\keywords{Mixed-traffic, Behaviour Modelling, Transferability, Calibration, Game Theory, Multi-agent Systems}

}

\begin{document}

\maketitle
\newcolumntype{M}{>{\centering\arraybackslash}m{2cm}}

\section{Introduction}
Shared space design principles \cite{emma2006shared} have been attracting notable attention over the past years as an alternative to traditional regulated traffic designs. Unlike the latter, \textit{shared space} design usually trades road signs, signals, and markings against higher levels of direct interaction between mixed traffic participants (e.g. cars, bikes, pedestrians); this interaction is largely guided by social protocols and informal rules. 

Numerous European cities reconstruct their city centres by adopting shared space principles; some examples are the Laweiplein intersection in the Dutch town Drachten, Skvallertorget in Norrk{\"o}ping, and Kensington High Street in London \cite{hamilton2008shared}. Increasing demands and in particular the absence of explicit traffic regulations and thereby caused vagueness makes it critical to investigate and assess safety issues in shared spaces during the planning phase.

Compared to regulated traffic designs, road users in shared spaces interact more frequently to coordinate their trajectories, to avoid and to resolve conflicts. Here, we define the notion of \textit{conflict} in accordance with \cite{gettman2003surrogate} as ``an observable situation in which two or more road users approach each other in time and space to such an extent that there is a risk of collision if their movements remain unchanged''.  Realistically modelling, simulating, and analysing the motion behaviour of mixed traffic users including interactions for conflict  avoidance or resolution can reproduce the operation of mixed-traffic environment, which can be useful for measuring traffic safety and efficiency (average road user delays and road capacity) of that environment \cite{danaf2020pedestrian}. \textbf{Note:} In this paper, conflict and interaction are used interchangeably.

Understanding the way heterogeneous road users make decisions and behave while interacting with others is far from trivial. Each interaction is the result of complex human decision-making processes, depends on the transport mode, the situation dynamics and  many individual factors (e.g. age, gender, or driving experience \cite{zheng2017driver}). 
Therefore, modelling mixed traffic interactions is challenging; to the best of our knowledge, until now, only few works investigate modelling and simulation of shared spaces. In related research on shared spaces modelling (see Section~\ref{sec:relatedwork}), we find two types of approaches: (1) extensions of the social force model (SFM) \cite{helbing1995social}, a physics-based model of pedestrian dynamics, to deal with heterogeneous physical actors \cite{schonauer2017microscopic,anvari2015modelling,rinke2017multi}; and (2) Cellular Automata (CA) models \cite{lan2005inhomogeneous,zhang2007modeling,bandini2017collision}, which are also used to model movement behaviour of heterogeneous road users in case there are clear rules for traffic regulations -- which is not the case in most shared space environments. 
 
The aforementioned approaches can well represent single bilateral conflicts, which means, for any road user and any point in time, only a single explicit conflict with one other user can be handled; however, in realistic shared space scenario, we often encounter the existence of multiple conflicts between different road users and groups (\cite{cheng2019pedestrian}). While the SFM provides some implicit capability to handle implicit conflicts using addition of forces, it does not cover explicit interaction, e.g. based on signalling or other communication. In our previous work, we combined SFM with a game theoretic model to also capture multiple conflicts \cite{johora2018modeling}.

However, previous research (including ours) has so far failed to address the \textit{transferability} of their simulation models. By this, we mean the ability of a model to reproduce realistic behaviour in environmental settings that are different from one(s) the model has been constructed for, with no or only little modification and adaptation.

In this paper, following a review of related work in Section~\ref{sec:relatedwork}, we describe the architecture of our Game-Theoretic Social Force Model (GSFM) for shared spaces (Section~\ref{sec:method}). Section~\ref{sec:calibration-methodology} contains the first contribution of this paper: we propose a generic method for calibrating our model for a set of given scenarios and illustrate its use for a shared space environment in Germany. The core of this method is a feature selection process used to find the features which have the highest influence on human decision-making in complex traffic situations. In  Section~\ref{sec:transferability}, we describe the second contribution, which is an investigation into transferability of shared space models. We define criteria for transferability of models, and present a case study in which we analyse and evaluate transferability of the model constructed based on the German dataset to a shared space environment in China, which is different in terms of culture (norms), traffic conditions, and spatial layout.

The results of this case study indicate that there are considerable differences in model characteristics in different cultural areas; however, by adding some social norm knowledge, we find that our initial model can be adapted to the new environment with reasonable effort, achieving satisfactory performance.


\section{Related Work}
\label{sec:relatedwork}
Previous works on modelling mixed traffic interactions mostly use rule-based models (e.g. Cellular Automata (CA) \cite{bandini2017approach}), or force-based models, most notably the Social Force Model (SFM) \cite{helbing1995social}.

In CA models, the environment consists of identical discrete cells and the movement of road users is conducted by a set of predefined rules. CA models have been used to model interaction among homogeneous road users e.g. pedestrians \cite{burstedde2001simulation,bandini2017approach}, cars \cite{nagel1992cellular,chai2015fuzzy} and also to model mixed traffic such as in \cite{lan2005inhomogeneous}, for modelling car-following and lane-changing behaviours of cars and motorcycles or in \cite{zhang2007modeling}, to model pedestrians-to-cars interactions at crosswalks or in \cite{chen2018evaluating}, for modelling the movement behaviours of bicycles and vehicles to analyse the influence of bicycle-to–vehicle conflicts on traffic delay.  
In the classical SFM, different motivations and interactions of road users are described by differential equations adding up a set of simple attractive and repelling forces influencing movement of a road user at a specific place and time. 
SFM was originally proposed for modelling pedestrian dynamics \cite{chen2018social,asano2010microscopic,johora2017dynamic}. 
It has been extended for modelling movement of vehicles, adding vehicle influence on pedestrians as separate forces \cite{yang2018social,zeng2014modified}, or by adding new forces for short-range conflicts and rule-based constraints for long-range conflicts \cite{anvari2015modelling}. \cite{rinke2017multi} extends SFM for modelling both vehicles and bicycles by using long-range collision avoidance mechanisms.

In simple scenarios, both CA-based and SFM-based models work well. While decision-theoretic models such as probabilistic and game-theoretic models are more capable to capture complex human decision-making process \cite{helbing1995social}. They can be used for handling complex conflict scenarios where agents need to choose an action among different alternatives to maximise their performance. In \cite{pascucci2018should}, Pascucci et al.~used a logit model for handling complex conflict scenarios. In a logit model, each decision maker chooses an action based on present data but without considering what others might do. In \cite{fujii2017agent}, Fujii et al. used a discrete choice model to capture pedestrian's decision making in pedestrian-to pedestrian interaction.
Game-theoretic models have been frequently used to illustrate the decision making processes of road users during interactions, for example, to describe pedestrian-to-car interaction in a shared space, merging-give way interaction of cars, lane-changing interaction of cars, or cyclist-car driver interaction at zebra crossings in \cite{schonauer2017microscopic,johora2018modeling, lutteken2016using} and \cite{bjornskau2017zebra}, respectively. In a non-cooperative game-theoretic model, each decision maker makes a decision by predicting others' decision, similar to real-world road users \cite{bjornskau2017zebra}. 

\section{Agent-Based Simulation Model}
\label{sec:method}
\begin{figure}
	\centering
	\includegraphics[width=3in,height=2.8in]{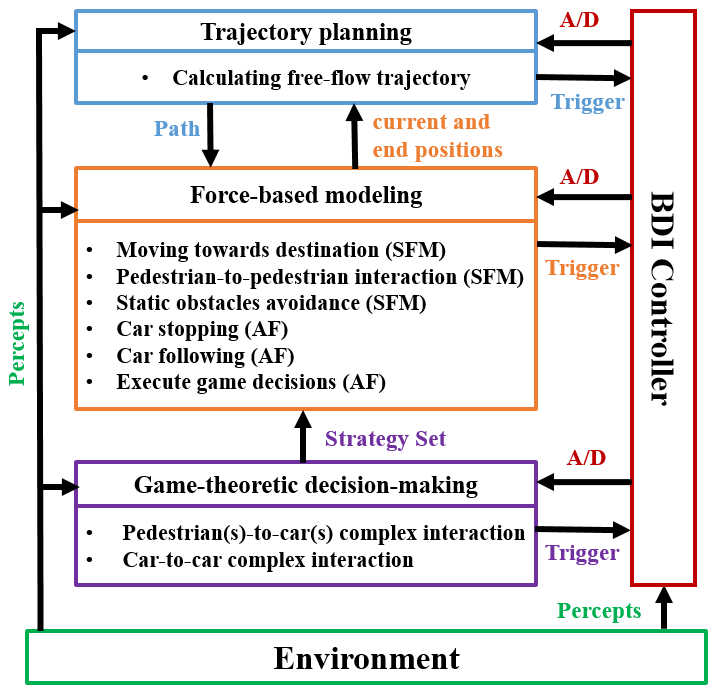}
	\caption[Conceptual model of motion behaviours of mixed traffic]{Conceptual model of motion behaviours of mixed traffic. Here, \textbf{AF} means added force to classical SFM and \textbf{A/D} denotes activation/deactivation of a module.}
	\label{fig:architecture}
	\vspace{-3.5mm}
\end{figure}

We have been developing agent-based model, named Game-Theoretic Social Force Model (GSFM), the initial version of this being described in \cite{johora2018modeling}, 
to represent and execute the movement behaviour of pedestrians and cars. In this section, we give a short overview of the principles underlying GSFM. In the GSFM architecture (see Fig.~\ref{fig:architecture}), the movement of each road user (in the following denoted as \textit{agent}) is handled by three  
interacting modules covering different functionality: \textit{trajectory planning}, \textit{force-based modelling}, and \textit{game-theoretic decision-making}. GSFM is implemented on a BDI (\textbf{B}elief, \textbf{D}esire, \textbf{I}ntention) platform, LightJason \cite{aschermann2016lightjason}, which allows to flexibly and declaratively design and explain the workflow logic of GSFM. Based on the current situation of the environment, the \textit{BDI controller} selects the module to be activated, which then notifies the controller upon completion of its task.

\textbf{The trajectory planning module} is responsible for computing free-flow trajectories for all agents by considering static obstacles such as boundaries, benches, status or trees in the shared space environment. We transform the simulation environment into a visibility graph \cite{koefoed2012representations} to perform the A* algorithm \cite{millington2009artificial} for trajectory planning.

\textit{The force-based modelling} and \textit{the game-theoretic decision-making} modules are in charge of handling interactions among road users. These interactions are categorised into two categories based on the observation of the shared space video data and on Helbing's classification of road agents' behaviour  \cite{helbing1995social}: simple interaction (percept $\rightarrow$ act) and complex interaction (percept $\rightarrow$ choose an action from different alternatives $\rightarrow$ act). 
These interactions can also be classified based on the number and types of road users involved namely, pedestrian-to-pedestrian, car-to-car, single or multiple pedestrian(s)-to-cars, and multiple pedestrian(s)-to-car interactions, which, also fall into the previous category. 

\textbf{The force-based module} is the operational module i.e. the physical execution of road users' movement is handled here. This module captures also simple interactions between road users. It uses the classical SFM to model each agent's driving force towards destination ({$\vec{D}_{i}^o$}), 
repulsive force from static obstacle ($\vec{I}_{i W}$) and from other road users ($\vec{I}_{i j}$). Here, $\vec{D}_{i}^o$ = $\frac{\vec{v^*}_i(t) - \vec{v}_i(t)}{\tau}$ for a relaxation time $\tau$ and $\vec{v^*}_i(t)$ and $\vec{v}_i(t)$ are the desired and current velocities of $i$ respectively, $\vec{I}_{i j}$ = $V_{i j}^o\exp\bigg[\frac{- \vec{d}_{i j}(t)}\sigma\bigg] \hat{n}_{i j} F_{i j}$ 
and 
$\vec{I}_{i W}$ =  $U_{i W}^o\exp\bigg[\frac{- \vec{d}_{i W}(t)}{R}\bigg] \hat{n}_{i W}$, where $V_{i j}^o$ and $U_{i W}^o$ denote the interaction strengths, and $\sigma$ and $R$ indicate the range of these repulsive interactions, $\vec{d}_{i j}(t)$ and $\vec{d}_{i W}(t)$ are the distances from $i$ to $j$, or $i$ to $W$ at a specific time, $\hat{n}_{i j}$ and $\hat{n}_{i W}$ denote the normalised vectors. $F_{i j}$ = $\lambda + (1 - \lambda)\frac{1+ \cos{\upvarphi_{i j}}}{2}$ represents the fact that human are mostly influenced by the objects which can be captured within their field of view \cite{johansson2008specification}. Here, $\lambda$
is a parameter which represents the strength of interactions from behind and $\upvarphi_{i j}$ denotes the angle between $i$ and $j$.
In GSFM, SFM is extended to capture car following interaction ($\vec{I}_\text{following}$) and pedestrian-to-vehicle reactive interaction ($\vec{I}_\text{stopping}$). 
If $d_{ij}$ $\geq$ $D_{min}$, $\vec{I}_\text{following}$ = $\hat{n}_{p x_j}$, $i$ maintains the direction of movement towards $p = \vec{x}_i(t) + \hat{v}_j(t) \cdot D_{min}$, otherwise, $i$ decelerates. Here, $D_{min}$ is the minimum vehicle distance, and $d_{ij}$ is the distance between $i$ and $j$ (leader car).
$\vec{I}_\text{stopping}$ happens only if pedestrian(s) have already initiated walking in front to the vehicle. Then the vehicle decelerates to let the pedestrian(s) pass.

\textbf{The game-theoretic module} is responsible for handling complex interactions i.e. pedestrian(s)-to-car(s) or car-to-car interaction. These interactions are modelled by Stackelberg game, a sequential leader-follower game where both leader and follower players try to maximise their utility: by leader choosing a strategy first considering all possible reactions of followers and the followers reacting based on leader's chosen strategy \cite{schonauer2017microscopic}. The game is solved (find the optimal strategy pair) by using the sub-game perfect Nash equilibrium (SPNE), depicted by Eq.~\eqref{eq:SPNE}. 
\vspace{-1.3mm}
\begin{equation}
\text{SPNE} = \{s_l\in S_l | max(u_l(s_l, Bs_f(s_l)))\}, \, \forall s_l\in S_l.
\label{eq:SPNE}
\end{equation}
\begin{equation}
Bs_f(s_l) = \{s_f\in S_f|max(u_f(s_f|s_l))\}.
\label{eq:Bs_f}
\end{equation}
The equation~\eqref{eq:Bs_f} represents the best answer from the follower. Here, $s_l$, $s_f$, $u_l$, $u_f$ and $S_l$, $S_f$ denote the strategies of leader and followers, utilities regarding the respective strategies and their strategy sets respectively. 
Each complex interaction is handled by an individual game and the games are independent on each other. For each game, the number of leaders is set to one but  the followers can be more, and the faster agent (i.e. car) is selected as the leader unless the situation involves more than one cars (e.g, pedestrian(s)-to-cars), then the one who detects the conflict first is set as leader. The way to calculate the payoff matrix of the game is discussed in Subsection  \ref{subsec:gamecalibration}. 
In GSFM, \textit{Continue}, \textit{Decelerate} and \textit{Deviate} (pedestrian only) are the possible actions for agents. 

\begin{itemize}
	\item Continue: Any pedestrian $\alpha$ crosses vehicle $\beta$ from the point $p_\alpha = {x}_\beta(t) + S_{A} * \overrightarrow{e}_\beta$, if  $line(x_\alpha(t), E_\alpha)$ intersects $line({x}_\beta(t) + S_{A} * \overrightarrow{e}_\beta, {x}_\beta(t) - \frac{S_A}{2} * \overrightarrow{e}_\beta)$, otherwise free-flow movement is continued. Here, $\overrightarrow{e}$ is the direction vector, $S_A$ denotes scaling factor, $x(t)$ and $E$ represent current and goal positions respectively. In case of vehicles, they always follow their free-flow movement.
	\item Decelerate: Road users decelerate and in the end stop (if necessary). For pedestrians, $\text{newSpeed}_{\alpha} = \frac{\text{Speed}_\alpha(t)}{2}$ and in case of vehicles, $\text{newSpeed}_{\beta} = \text{Speed}_{\beta}(t) - \text{decRate}$.\newline
	\\
	Here, 
	$\text{decRate} = \begin{cases}
    \frac{\text{Speed}_{\beta}(t)}{2}, \text{if } \text{distance}(\alpha, \beta) \leq D_{min}, \newline
    \\
    \frac{\text{Speed}_{\beta}^2}{\text{distance}(\alpha, \beta) - D_{min}}, \text{otherwise}.
   \end{cases}$ 
   \newline
   $D_{min}$ is the critical spatial distance.
    \item Deviate: A pedestrian $\alpha$ passes a vehicle $\beta$ from behind from a position $p_\alpha = {x}_\beta(t) - S_A*\overrightarrow{e}_\beta(t)$ (as long as $\beta$ stays in range of the field of view (FOV) of $\alpha$) and after that $\alpha$ resumes moving towards its original destination. 
\end{itemize}

Although these modules take control alternatively without following any sequence, 
at the start of the simulation, GSFM maintains a hierarchy among them: it starts with trajectory planning with the premise that agents plan their trajectories (at the granularity of landmarks) before they actually start moving. Once trajectories are obtained, the force-based module is activated to bring about the agents' physical movement.
Conflict recognition and classification are performed at regular intervals using Algorithm \ref{condetection}.
In case of complex conflict, the BDI controller activates the game-based module. Once the strategies are selected, the BDI controller activates the force module 
to execute the selected strategies. The BDI controller also prioritises different interactions based on their severity, such as, for cars, $\vec{I}_\text{stopping}$ takes priority over $\vec{I}_\text{game}$ and $\vec{I}_\text{game}$ gets precedence on simple interaction like car following. 

To sum up, the process of modelling the movement behaviour of any target agent $i$ at any time step $t$ in GSFM is presented in Eq.~\eqref{eq:PedSFM}--\eqref{eq:CarSFM}. 

\begin{equation}
\vspace{-3mm}
\text{Pedestrian:~}   
 \frac {d{\overrightarrow{v^t}_i}}{dt} =	\Big(\overrightarrow{D}_{i}^o + \Sigma\overrightarrow{I}_{i W} + \Sigma\overrightarrow{I}_{ij} \Big) \hspace{0.1cm} or \overrightarrow{I}_{\text{game}},
 \label{eq:PedSFM}
\end{equation}
\begin{equation}
\text{Car:~}
 \frac {d{\overrightarrow{v^t}_i}}{dt} =	\overrightarrow{D}_{i}^o \hspace{0.1cm} or \overrightarrow{I}_{\text{following}} \hspace{0.1cm} or \overrightarrow{I}_{\text{game}} \hspace{0.1cm} or \overrightarrow{I}_{\text{stopping}},
 \label{eq:CarSFM}
 \vspace{-3mm}
\end{equation}

\begin{algorithm*}
  \begin{algorithmic}[1]
    \Function{Conflict Recognition}{$setofCars,setofPedestrians$}  \Comment{$V_R \gets$ View Range, $S_C \gets$ Scaling Factor} 
    \State $\vars{conflictSet} \gets  \{\}$ \Comment{$x(t) \gets$ Current Position, $G \gets$ Goal, $\vec{e} \gets$ Moving Direction}
    \State $\vars{setofAll} \gets setofCars \cup setofPedestrians$ \Comment{ $\hat{n} \gets$ Normalized Vector, $D_{min} \gets$ Critical Spatial Distance}
    \ForEach{$i \in setofCars$} \Comment{\parbox[t]{7.1cm}{\raggedright$i.getPreviousConflict() \gets $ get competitive users of i from previously detected but active conflicts of i }}
     \State $\vars{competitiveCars} \gets  \{\}$, $\vars{competitivePeds} \gets  \{\}$ \Comment{$j.getConflict() \gets $ get competitive users of j}
      \If{$i.inIntersectionZone()$} 
        \ForEach{$j \in \vars{setofAll} \hspace{0.1cm}\textbf{and}\hspace{0.1cm} j \neq i  \hspace{0.1cm}\textbf{and} \hspace{0.1cm} \nexists \hspace{0.05cm} j\in  i.getPreviousConflict()  \hspace{0.07cm}\textbf{and} \hspace{0.07cm} \nexists \hspace{0.05cm} i\in  j.getConflict()$}
          \If{ $distance(i,j) \leq V_R \hspace{0.1cm}\textbf{and}\hspace{0.1cm} 
              ( ( \exists \hspace{0.05cm} j\in setofCars \hspace{0.1cm}\textbf{and}\hspace{0.1cm} (\theta_{\vec{e}_i\hat{n}_{ji}} \leq 90 \degree \hspace{0.2cm} \Or \hspace{0.2cm} \theta_{\vec{e}_i\hat{n}_{ji}} \geq 270 \degree)\hspace{0.2cm}) \hspace{0.1cm} \Or \newline
             \hspace*{2.5em}( \exists \hspace{0.05cm} j\in setofPedestrians \hspace{0.1cm}\textbf{and}\hspace{0.1cm} (\theta_{\vec{e}_i\hat{n}_{ji}} \leq 113 \degree \hspace{0.1cm} \Or \hspace{0.2cm} \theta_{\vec{e}_i\hat{n}_{ji}} \geq 247 \degree) \hspace{0.1cm}) \hspace{0.1cm}$)} 
                \State $\vars{$predictedPosition_i$}$ $\gets$ $x_i(t)$ + $S_C$ * i.maxSpeed() * $\vec{e}_i$
                \State $\vars{$predictedPosition_j$}$ $\gets$ $x_j(t)$ + $S_C$ * j.maxSpeed() * $\vec{e}_j$
                \If{distance($\vars{$predictedPosition_i$}$, $\vars{$predictedPosition_j$}$) $\leq$ $D_{min}$}
                  \If{$\exists \hspace{0.05cm} j \in setofCars$}
                    \hspace{0.1cm} $\vars{competitiveCars} \cup \{j\} $ 
                  \Else
                    \hspace{0.1cm} $\vars{competitivePeds} \cup \{j\} $ 
                  \EndIf
                 \EndIf 
          \EndIf
        \EndFor
      \Else
        \ForEach{$j \in setofPedestrians \hspace{0.1cm}\textbf{and} \hspace{0.1cm} \nexists \hspace{0.05cm} j\in  i.getPreviousConflict()  \hspace{0.1cm}$}
          \If{ $distance(i,j) \leq V_R \hspace{0.1cm}\textbf{and}\hspace{0.1cm} (\theta_{\vec{e}_i\hat{n}_{ji}} \leq 113 \degree \hspace{0.2cm} \Or \hspace{0.2cm} \theta_{\vec{e}_i\hat{n}_{ji}} \geq 247 \degree$)}
            \State $backPosition_j \gets$ $x_j(t)$ - j.diameter() *  $\vec{e}_j$
            \If{lineSegmentIntersect( $backPosition_j$, $G_j$, $x_i(t)$, $G_i$ )}
            \hspace{0.1cm} $\vars{competitivePeds} \cup \{j\} $ 
            \EndIf
          \EndIf
        \EndFor
      \EndIf
      \State  $\vars{conflictSet} \cup \Call{Conflict Classification}{$i$, $\vars{competitivePeds}$, $\vars{competitiveCars}$,$\vars{setofCars}$}$
    \EndFor
    \State \Return $\vars{conflictSet}$
    \EndFunction
    
    \Statex
    
    \Function{Conflict Classification}{$u$,$p$,$c$,$cars$} \funclabel{alg:b}
    \State  $\vars{competitiveUsers} \gets  \{\}$,  $\vars{mC} \gets  \{\}$
    \If{ $p \neq \emptyset \hspace{0.1cm}\textbf{and}\hspace{0.1cm} c \neq \emptyset$ }
      \hspace{0.1cm} \Return \{u, \vars{competitiveUsers} $\cup$ p $\cup$ c, \vars{Pedestrian(s)-to-Cars}\}
    \ElsIf{ $p == \emptyset \hspace{0.1cm}\textbf{and}\hspace{0.1cm} c \neq \emptyset$}
      \hspace{0.1cm} \Return \{u, \vars{competitiveUsers} $\cup$ c, \vars{Car-to-Car}\}
    \ElsIf{$u.inIntersectionZone()$ \hspace{0.1cm}\textbf{and}\hspace{0.1cm} $p \neq \emptyset \hspace{0.1cm}\textbf{and}\hspace{0.1cm} c == \emptyset$ }
         \State \Return \{u, \vars{competitiveUsers} $\cup$ p, \vars{Pedestrian(s)-to-Car}\}
    \ElsIf{$u.inRoadZone() \hspace{0.1cm}\textbf{and}\hspace{0.1cm} p \neq \emptyset$ }
         \State $\forall x \in cars\textbf{:} \hspace{0.1cm} \textbf{if} \hspace{0.1cm}( x.nearestCompetitor()  ==  u.nearestCompetitor()) \hspace{0.1cm} \textbf{then} \hspace{0.1cm} mC \hspace{0.05cm}\cup \hspace{0.05cm}\{x\},\hspace{0.1cm} x.getConflict() \gets \{\} $  
         \If{ mC $\neq \emptyset$} \hspace{0.1cm} \Return \{u, \vars{competitiveUsers} $\cup$ p $\cup$ mC, \vars{Pedestrian(s)-to-Cars}\}
         \Else \hspace{0.1cm}\Return \{u, \vars{competitiveUsers} $\cup$ p, \vars{Pedestrian(s)-to-Car}\}
         \EndIf
    \Else
       \hspace{0.1cm}\Return \{u, \vars{competitiveUsers}, \vars{No New Conflict}\}
    \EndIf
    \EndFunction

  \end{algorithmic}
  \caption{Conflict Recognition Algorithm}\label{condetection}
\end{algorithm*}

\section{Calibration Methodology}
\label{sec:calibration-methodology}
The GSFM model has a large number of parameters, which can be classified into three categories: (1) parameters of the SFM interactions, e.g. range of repulsive interaction between pedestrians, (2) safety measurements, e.g.  pedestrian-to-car safety distance, and (3) parameters defining the payoff matrices for game playing. 

\begin{figure}[htbp]
\vspace{-1mm}
	\centering
	\includegraphics[width=7.25cm]{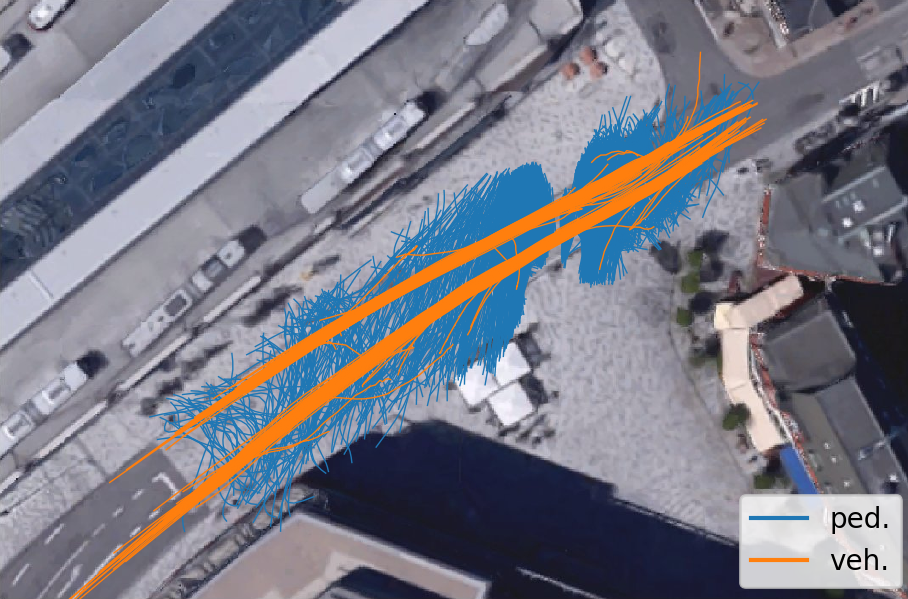}
	\caption{Mixed trajectories from a shared space in Hamburg.}
	\label{fig:hamburgdataset}
\end{figure}

To calibrate these parameters and validate the performance of the proposed model, we extract 104 real--world interaction scenarios from the Hamburg Bergedorf station dataset (HBS) in the German City Hamburg \cite{pascucci2017discrete}. 
This dataset contains trajectories of 1115 pedestrians and 331 vehicles from one large video clip. 
 The layout of this shared space is a street with pedestrian crossing from both sides, as visualised in Figure \ref{fig:hamburgdataset}. At each time step, a road user's position is given by a 2D vector in the pixel coordinate system and the pixel-to-meter conversion scale is also given with the dataset. Each time step in HBS is equal to 0.5 seconds. Extracted scenarios include pedestrian-to-car interactions with or without car following interactions, multi-user (MUC) i.e. conflict among multiple road users, where all users in the conflict are co-dependent e.g. pedestrians-to-car interaction and multiple conflicts (MC) where all users are not co-dependent like pedestrian(s)-to-cars road crossing scenario. We divide the scenarios in two parts, one for calibration (66\%) and another part for validation (34\%). 

The criteria we use for estimating the performance of our model in both calibration (see Section \ref{sec:calibration-methodology}) and evaluation processes (\ref{sec:transferability}) are: 
\begin{enumerate}
\item the \textbf{displacement error}, i.e., the average Euclidean distance error (ADE) that measures the aligned error for each step and report the value averaged over the path.
\item the \textbf{speed deviation} to measure the pairwise speed difference of simulated and real speed of each road user over all time steps.
\item \textbf{error in decision-making} to measure the difference between the simulated decision and the decision taken by the real road users to handle the same situation (use only for calibration).
\end{enumerate}

We calibrate the parameters of our model in two steps: We start with calibrating the SFM and safety parameters using a genetic algorithm (see Subsection~\ref{subsec:genalg}; then, we select the important features for payoff estimation using backward elimination and optimise the value of the parameters related to these features using a genetic algorithm  (Subsection~\ref{subsec:select-features}). More details are given in the Subsections \ref{subsec:sfmcalibration} and \ref{subsec:gamecalibration}. Subsection~\ref{subsec:calibrationresults} shows the results of applying the method to a real-world data set from Germany.

\subsection{Genetic Algorithm}
\label{subsec:genalg}
\begin{figure}
	\centering
	\includegraphics[width=3.3in,height=2.3in]{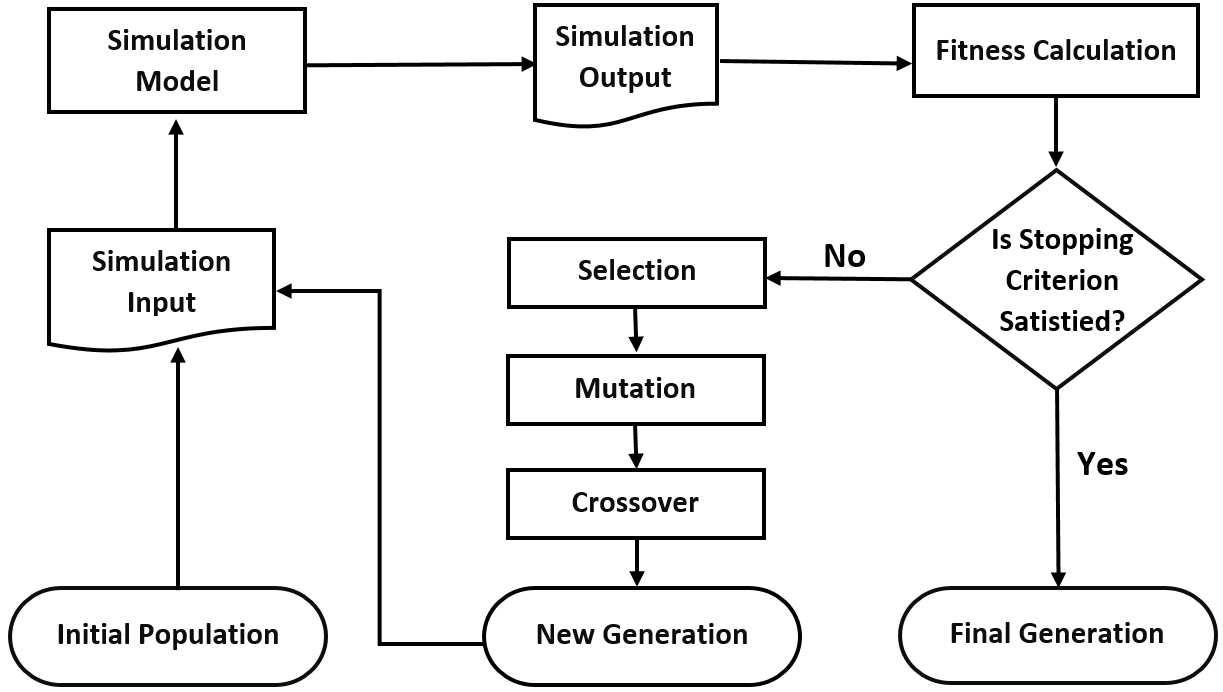}
	\caption{Parameter calibration workflow (adapted from~\cite{cunha2009genetic}).}
	\label{fig:genalgo}
	\vspace{-5mm}
\end{figure}

Genetic algorithms \cite{zames1981genetic}  reflect natural processes of evolution by selecting the fittest individuals of the current population in order to produce offspring for the next generation. These algorithms are widely used to address optimisation problems, and are also applied to  calibrating model parameters 
\cite{amirjamshidi2019multi,cunha2009genetic}. The main properties of a genetic algorithm are as follows:
\begin{itemize}
\item \textbf{Population:} The algorithm starts with a set of individuals, called a \textit{population}. Each individual, known as \textit{chromosome}, is a candidate solution to the problem: in our case, the values of the chosen parameters for calibration. Each parameter in a chromosome is called \textit{gene}. Genes are appended into a string to form a chromosome.
\item \textbf{Fitness Function}: Calculates the fitness and gives a score to each candidate solution. 
\item \textbf{Selection operator}: Chooses the individuals of the population as parents, randomly or by using fitness as probability, for the creation of offspring.   
\item \textbf{Crossover}: The chosen parents in the selection phase exchange their genes among themselves until the crossover point is reached. A crossover point is chosen by e.g. arithmetic operations or random choice.
\item \textbf{Mutation}: To keep population diversity and prevent premature convergence, this process randomly modifies the genes of the offspring.
\end{itemize}
The workflow between these processes to calibrate the parameters of our simulation model are visualised in Figure~\ref{fig:genalgo}. The algorithm starts with a random initial population. Each chromosome, i.e. set of parameters, is fed into the simulation model to obtain outputs, which are later compared with real-world data to calculate the fitness score of the chromosome. The types of required simulation output and the fitness function depends on the types of parameters to calibrate (see section \ref{subsec:sfmcalibration} and \ref{subsec:gamecalibration}). An offspring population is generated based on the selection, crossover and mutation processes. Feeding into the simulation model, the fitness calculation and the generation of new offspring continues until a specific stopping criteria is fulfilled. 

\subsection{Feature Selection}
\label{subsec:select-features}
When modelling real--world situations, selecting most relevant features from many potential variables
 is an important task to decrease the complexity and improve the performance of the model. Backward elimination is one of the simple but widely used feature selection methods \cite{garcia2014process}. The steps of this method are as follows:
\begin{itemize}
    \item[1.] Select a significance level ($p$-value, e.g. for this paper: 0.09). A $p$-value is a statistical measure that is used to determine whether the null hypothesis is true or not. Our null hypothesis is that the chosen variables for the model have no influence on the results.
    \item[2.] Fit the model with all variables.
    \item[3.] Select the variable with the highest $p$-value. If its $p$-value is greater than the preset value, go to step 4; else the null hypothesis will be rejected. Which means the remaining set of features are relevant to the model, so the process will terminate.
    \item[4.] Eliminate this feature.
    \item[5.] Fit the model with the rest of the features, and go back to step 3.
\end{itemize}

\subsection{Calibration of SFM and Safety Parameters}
\label{subsec:sfmcalibration}
The list of the SFM and safety parameters that need to be calibrated is given in Table \ref{table:sfmparameters}. Here, $PP$, $PC$, and $CC$ denote pedestrian-to-pedestrian,  pedestrian-to-car, and car-to-car interactions, respectively.
\begin{table*}
\centering
\setlength{\tabcolsep}{8.5pt}
\renewcommand{\arraystretch}{1.5}
\caption{The list of the SFM and safety parameters with their calibrated values}
\begin{tabular}{ |p{1.5cm}||p{2.5cm}|p{6.4cm}|p{1cm}|p{2.7cm}|  }
 \hline
 \textbf{Symbol} & \textbf{Type} & \textbf{Description} & \textbf{Unit} & \textbf{Value} \\

 \hline

 $V_{\alpha\beta}^o$   & SFM    &Interaction strength for other user&  $m^2s^{-2}$ & 1.4 ($PP$), 10.0 ($PC$)\\
 \hline
 $U_{\alpha B}^o$ &   SFM  & Interaction strength for obstacle   &$m^2s^{-2}$ & 10.0\\
  \hline
 $\sigma$ &   SFM  & Range of repulsive interaction for other  &$m$ & 0.4 ( $PP$), 0.2 ($PC$)\\
   \hline
 $R$ &   SFM  & Range of repulsive interaction for obstacle   &$m$ & 0.2\\
  \hline
 $\lambda$ &   SFM & Anisotropic parameter  & --- & 0.2\\
  \hline
 $D_{min}$ &   Safety measure  & Critical spatial distance   &$m$ & 8.0 ($PC$ \& $CC$)\\
  \hline
 $S_A$ &   Safety measure  & Scaling factor for Continue \& deviate actions   &$m$ & 7.0\\
  \hline
 $V_R$ &   Safety measure  & Range of view   &$m$ & 18.4\\
 \hline
 $S_C$ &   Safety measure  & Scaling factor for Conflict recognition  & --- & 9.0\\
\hline
\end{tabular}
\label{table:sfmparameters}
\end{table*}

For calibrating these twelve parameters (including six single-valued and three parameters with multiple values), we use the genetic algorithm, described in Subsection \ref{sec:calibration-methodology}.\ref{subsec:genalg}. The algorithm starts with a randomly generated population. Each input chromosome to GSFM is formed of twelve genes (for twelve parameters); GSFM simulates all training scenarios from the dataset (HBS) and gives the simulated position of each road users of each scenarios as the output. The simulated and real positions of road users are then compared to calculate and assign the fitness score to the respective chromosome using the fitness function denoted in Equation \ref{eq:sfmfitnessfunction}. The last column of Table \ref{table:sfmparameters} shows the values of the respective parameters.

\begin{equation}
   f_{score} = 
   \Bigg( \sum^{E}_\textit{e} \bigg(\sum^{U}_\textit{u} \Big( \sum^{T}_\textit{t}\abs{\vec{P}^{real}_u (\textit{t}) \hspace{0.02cm} - \hspace{0.02cm} \vec{P}^{Simulated}_u (\textit{t})}\Big)\big/T \bigg) \Big/ U \Bigg) \bigg/E
   \label{eq:sfmfitnessfunction}
\end{equation}

 \subsection{Calibration of Game Parameters} 
\label{subsec:gamecalibration}

To calculate the payoff matrix of the game, firstly, all actions of the players are ordinally valued with the assumption that they prefer to reach their destination safely and quickly. Secondly, nine relevant observable features are considered to capture courtesy behaviours and situation dynamics. 
Let, $\alpha$ be a road user who needs to interact with another user $\beta$, then the feature are the following:
\begin{itemize}
	\item\textbf{OwnSpeed}: 
	\[ \begin{dcases}
S_{current}, & if \hspace{0.05cm} \alpha \hspace{0.05cm}is \hspace{0.05cm}a \hspace{0.05cm}car \\
	1, & if \hspace{0.05cm} \alpha \hspace{0.05cm}is \hspace{0.05cm}a \hspace{0.05cm}pedestrian \And S_{current} > S_{high}  \\
	0, & \text{otherwise}
	\end{dcases}\]
	\item\textbf{CompetitorSpeed}: has value $1$, if current speed of $\beta$, $S_{current}$ $<$ $S_{normal}$, otherwise 0. 
	\item \textbf{NOAI}: the number of active interactions of $\alpha$ as a car.
	\item \textbf{CarStopped}: has value $1$ if $\alpha$ ( as a car) already stopping to give way to another user $\beta'$, otherwise 0. 
	\item \textbf{CarFollowing}: has value $1$ if $\alpha$ is a car driver following another car $\beta'$,  otherwise 0.
	\item \textbf{Angle}: 
\[
\begin{dcases}
    8,& if (\theta_{\vec{e}_\beta\hat{n}_{\alpha\beta}}< 16\degree \And\theta_{\vec{e}_\beta\hat{n}_{\alpha\beta}}\geq 0\degree) \hspace{0.03cm} \Or \hspace{0.03cm} \theta_{\vec{e}_\beta\hat{n}_{\alpha\beta}}>344\\
    7,& if (\theta_{\vec{e}_\beta\hat{n}_{\alpha\beta}}\leq 42\degree \And\geq 16\degree) \hspace{0.02cm} \Or \hspace{0.02cm}  (\theta_{\vec{e}_\beta\hat{n}_{\alpha\beta}}\leq 344\degree \And \geq 318\degree)\\
    6,& if (\theta_{\vec{e}_\beta\hat{n}_{\alpha\beta}}\leq 65\degree \And > 42\degree) \hspace{0.02cm} \Or \hspace{0.02cm} \notag (\theta_{\vec{e}_\beta\hat{n}_{\alpha\beta}}< 318\degree \And \geq 295\degree)\\
    5,& if \hspace{0.02cm} (\theta_{\vec{e}_\beta\hat{n}_{\alpha\beta}}\leq 90\degree \And > 65\degree) \hspace{0.02cm} \Or \hspace{0.02cm} \notag (\theta_{\vec{e}_\beta\hat{n}_{\alpha\beta}}< 295\degree \And \geq 270\degree)\\
    1,              & \text{otherwise}
\end{dcases}
\]
	\item \textbf{CarFollowed}: has value $1$ if $\alpha$ is a car driver followed by another car $\beta'$, otherwise 0.
	\item \textbf{MinDist}: has value $D_{min}$ - distance($\alpha$, $\beta$), if distance($\alpha$, $\beta$) $<$ $D_{min}$ i.e. $\alpha$ (car) is unable to stop, otherwise 0.
	\item \textbf{GivewayNr}: number of give way  of $\alpha$ as a car.
\end{itemize}
To select the most relevant features among the above-mentioned features for pedestrian-to-car interactions from the perspective of both pedestrian and car,  we perform the backward elimination process, described in Subsection \ref{subsec:select-features}. To compute the p-value of each features, we choose a multinomial logit model because of its categorical structure of response;
three discrete outcomes for pedestrian and two for car, as visualised in Table \ref{table:beforebackwardcar}--\ref{table:afterbackwardped}.

\begin{table}[!htbp]
\centering
\setlength{\tabcolsep}{8.5pt}
\renewcommand{\arraystretch}{1.5}
\caption{Before backward elimination (\textbf{Car})}
\begin{tabular}{ |p{2.5cm}||p{1.5cm}|p{1.5cm}|p{1cm}|  }
 \hline
 \textbf{Variable} & \textbf{Coefficient} & \textbf{Standard error} & \textbf{P-value} \\
 \hline
 OwnSpeed&	-0.6371&	0.1178&	0.0000\\
 \hline
CompetitorSpeed&0.5841	&0.2558&	0.0224\\
  \hline
NOAI&	0.2352&	0.1087&	0.0305\\
   \hline
CarStopped&	1.7215	&0.3960	&0.0000\\
  \hline
 Angle&	0.0755&	0.0677&	0.2654\\
  \hline
 CarFollowing&	0.0627&	0.2533&	0.8044\\
  \hline
MinDist&	-0.0548&	0.0244&	0.0246\\
  \hline
GivewayNr&	-0.3577	&0.1251	&0.0043\\
  \hline
\end{tabular}
\label{table:beforebackwardcar}
\end{table}

\begin{table}[htbp]
\centering
\setlength{\tabcolsep}{8.5pt}
\renewcommand{\arraystretch}{1.5}
\caption{After backward elimination (\textbf{Car})}
\begin{tabular}{ |p{2.5cm}||p{1.5cm}|p{1.5cm}|p{1cm}|  }
 \hline
 \textbf{Variable} & \textbf{Coefficient} & \textbf{Standard error} & \textbf{P-value} \\
 \hline
 OwnSpeed&	-0.6363&	0.1177&	0.0000\\
 \hline
CompetitorSpeed&	0.5927&	0.2536&	0.0194\\
  \hline
NOAI&	0.2368&	0.1084&	0.0290\\
   \hline
CarStopped&	1.7116&	0.3936&	0.0000\\
  \hline
 Angle&	0.0690&	0.0626&	0.2698\\
  \hline
MinDist&	-0.0553&	0.0243&	0.0229\\
  \hline
GivewayNr&	-0.3569&	0.1250&	0.0043\\
  \hline
\end{tabular}
\label{table:afterbackwardcar}
\end{table}

\begin{table*}[htbp]
\centering
\setlength{\tabcolsep}{8pt}
\renewcommand{\arraystretch}{1.5}
\caption[Before backward elimination (\textbf{Pedestrian}) ]{Before backward elimination (\textbf{Pedestrian}) : taken \textit{accelerate} as the baseline strategy for the logit model}
\begin{tabular}{|c|c|c|c|c|c|c|}
 \cline{2-7}
\multicolumn{1}{c|}{}& \multicolumn{3}{|c}{Accelerate $\rightarrow$ Decelerate} & \multicolumn{3}{|c|}{Accelerate $\rightarrow$ Deviate}\\ \hline
\textit{Variable} &	Coefficient &	Standard error&	P-value  &	Coefficient &	Standard error&	P-value     \\
\hline 
\textit{OwnSpeed}&	0.5386&	0.1140&	0.0000&		0.2204&	0.1489&	0.1390\\
\hline
\textit{CompetitorSpeed}&	-2.3644	&0.6279&	0.0002&		-1.1938&	0.7315&	0.1027\\
\hline
\textit{CarStopped}&	-0.3278&	0.0525&	0.0000&		-0.0418&	0.0728&	0.5661\\
\hline
\textit{Angle}&	0.6576&	0.3031&	0.0300&		0.6796&	0.3970&	0.0869\\
\hline
\textit{CarFollowed}&	1.5395&	0.9734&	0.1137	&-0.9600&	0.7152&	0.1795 \\
\hline
\end{tabular}
\label{table:beforebackwardped}
\end{table*}

\begin{table*}[htbp]
\centering
\setlength{\tabcolsep}{8pt}
\renewcommand{\arraystretch}{1.5}
\caption[After backward elimination (\textbf{Pedestrian})]{After backward elimination (\textbf{Pedestrian}) : taken \textit{accelerate} as the baseline strategy for the logit model}
\begin{tabular}{|c|c|c|c|c|c|c|}
 \cline{2-7}
\multicolumn{1}{c|}{}& \multicolumn{3}{|c}{Accelerate $\rightarrow$ Decelerate} & \multicolumn{3}{|c|}{Accelerate $\rightarrow$ Deviate}\\ \hline
\textit{Variable} &	Coefficient &	Standard error&	P-value  &	Coefficient &	Standard error&	P-value     \\
\hline 
\textit{OwnSpeed}&	0.5429&	0.1125&	0.0000&		0.1175&	0.0699&	0.0925\\\hline 
\textit{CompetitorSpeed}&	-2.3320&	0.6261&	0.0002&	-1.3282&	0.6926	&0.0552\\\hline 
\textit{CarStopped}&	-0.3259&	0.0520&	0.0000&&&\\	\hline 
\textit{Angle}&	0.6805&	0.3016&	0.0241	&0.5662&	0.3257&	0.0822\\			
\hline
\end{tabular}
\label{table:afterbackwardped}
\end{table*}

\begin{figure*}[!htbp]
	\centering
	\subfloat[Pedestrian-to-Car Interaction]{\includegraphics[width=1.7in, height = 1.2in]{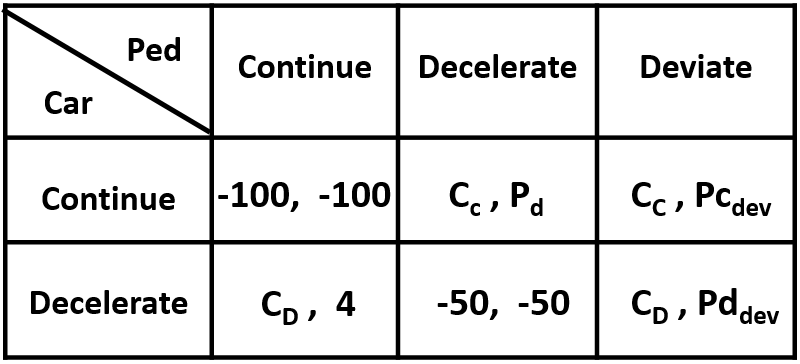}}
	\hfil
	\subfloat[Impacts of Situation Dynamics]{\includegraphics[width=5.3in, height = 1.2in]{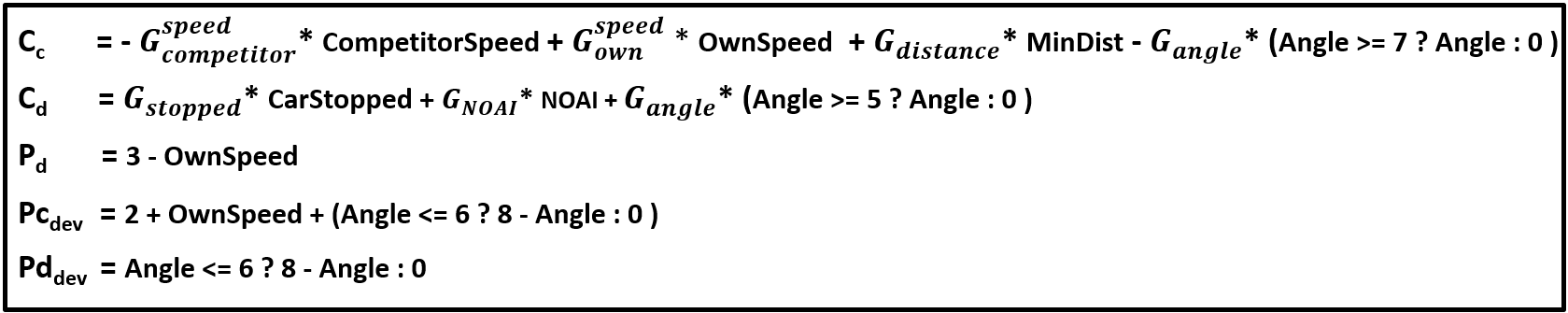}}
	\caption{The complete payoff matrices of pedestrian-to-car  interactions with all considered actions.}
	\label{gamematrix}
    \vspace{-0.75em}
\end{figure*}
The list of the game parameters with their calibrated values is given in Table \ref{table:gameparameters}.
For calibrating these parameters, we also use the genetic algorithm, given in Section \ref{subsec:genalg} but with a different fitness function, described in Equation \ref{eq:gamefitnessfuction}. For each training-scenario from the HBS dataset, the real and simulated decisions (i.e. accelerate/decelerate/deviate) of each involved road users, to handle that scenario, are compared to calculate the fitness score. The simulated decisions are computed using the payoff matrix in Figure \ref{gamematrix}, built on the selected features given in Table \ref{table:afterbackwardcar} and \ref{table:afterbackwardped}.

\begin{table}[htbp]
	\centering
	\setlength{\tabcolsep}{8.5pt}
	\renewcommand{\arraystretch}{1.5}
	\caption{The list of game parameters with calibrated values}
	\begin{tabular}{ |p{1.5cm}||p{0.7cm}|p{1.5cm}||p{0.7cm}|  }
		\hline
		\textbf{Symbol} & \textbf{Value} & \textbf{Symbol} & \textbf{Value} \\
		\hline
		$G^{own}_{speed}$   & 11 & $G_{noai}$  & 3\\
		\hline
		$G^{competitor}_{speed}$ & 11 & $G_{stopped}$  & 2 \\
		\hline
		$G_{angle}$ & 1 & $G_{distance}$ &  1 \\
		\hline
	\end{tabular}
	\label{table:gameparameters}
\end{table}

\begin{equation}
f_{score} = 
\sum^{E}_\textit{e} \Bigg(
\bigg(\sum^{U}_\textit{u}
\begin{cases}
1, & A^{real}_u \hspace{0.02cm} == \hspace{0.02cm} A^{Simulated}_u\\
-1, & \text{otherwise}
\end{cases}
\bigg)
\bigg/ U \Bigg) 
\Bigg/ E
\label{eq:gamefitnessfuction}
\end{equation}
 \textbf{Note:} Because of insufficient car-to-car complex interaction scenarios, we could not calibrate the game parameters for car-to-car interactions. Even though the \textit{Angle} feature for the decision making of cars has a higher p-value than our preset value, \textit{Angle} is used to calculate car's utility to maintain higher accuracy. Keeping counts on the number of times / frequency a car has been giving way in car-to-pedestrian interactions and considering this for calculating the payoff matrix is part of our future work.

\subsection{Calibration Results}
\label{subsec:calibrationresults}

\begin{figure*}[htbp]
	\centering
	\vspace{-1.15em}
	\subfloat[The decision matrix for cars on training-scenarios]{\includegraphics[width=1.67in]{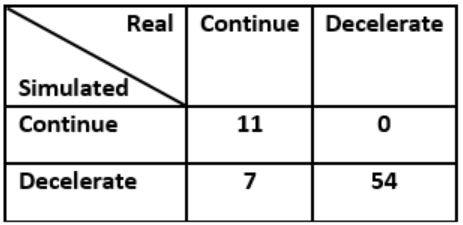}}
	\hfil
	\subfloat[The decision matrix for pedestrian on training-scenarios ]{\includegraphics[width=1.7in]{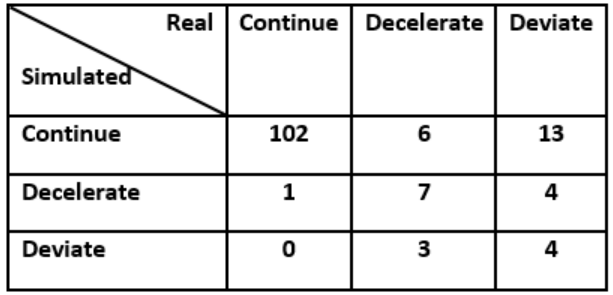}}
		\hfil
	\subfloat[The decision matrix for cars on testing-scenarios]{\includegraphics[width=1.67in]{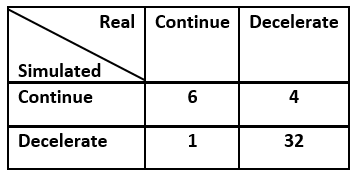}}
	\hfil
	\subfloat[The decision matrix for pedestrian on testing-scenarios]
	{\includegraphics[width=1.7in]{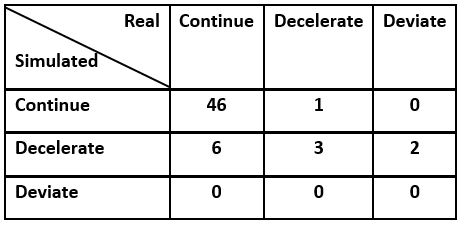}}
	\caption{The decision matrices of pedestrians and cars.}
	\label{decisionmatrix}
    \vspace{-0.75em}
\end{figure*}

The accuracy of our model in decision making is shown in Figure \ref{decisionmatrix}. The off-diagonal elements of the decision matrices denote the situations where the simulated decision differs from the observed one. The accuracy of 90.27\% (for cars) and 80.71\% (for pedestrians) on the training-scenarios and 88.37\% (for cars) and 84.48\% (for pedestrians) on the testing scenarios can be considered satisfactory, given the fact that the behaviours of road users may strongly depend on individual properties (such as age, gender) and conditions e.g. time pressure, which are currently not considered considered in our model.

\begin{figure}[htbp]
	\centering
	\subfloat[Trajectory Difference ]{\includegraphics[width=2.8in]{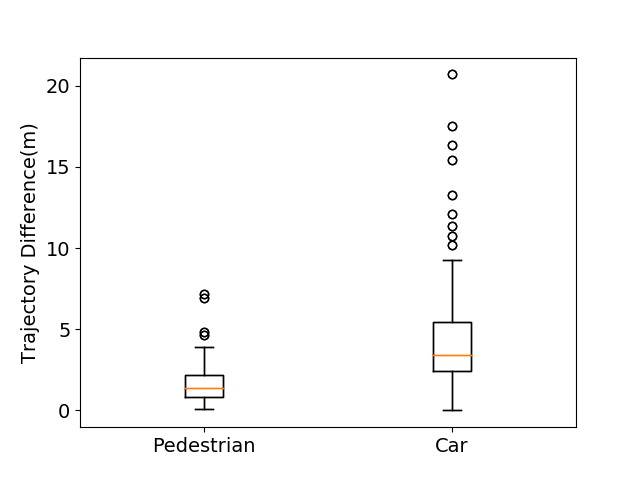}}
	\hfil
	\subfloat[Speed differences ]{\includegraphics[width=2.57in]{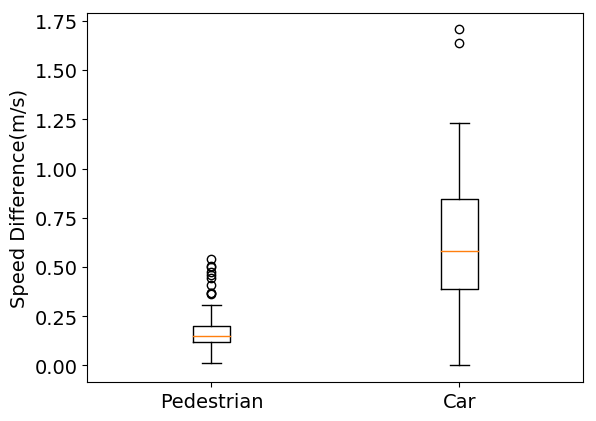}}
	\caption{The difference in trajectory and speed of real and simulated
pedestrians and cars of the HBS dataset.}
	\label{fig:QuantitativeHamburg}
\end{figure}

Figure \ref{fig:QuantitativeHamburg} visualises the deviation of real from simulated trajectories and speeds of all road users involved in 104 selected real scenarios from the HBS dataset.
The mean and standard deviation in the speed difference are: 0.6430 $ms^{-0.5}$ and 0.3179 $ms^{-0.5}$ (cars);  0.1767 $ms^{-0.5}$ and 0.1051 $ms^{-0.5}$ (pedestrians) and in the trajectory difference are:  4.6274 $m$ and 3.720 $m$ (cars); 1.7013 $m$ and 1.2907 $m$ (pedestrians).

\section{Evaluation of Transferability} 
\label{sec:transferability}
\begin{figure}[htbp]
	\centering
	\subfloat[Roundabout]{\includegraphics[width=1.9in]{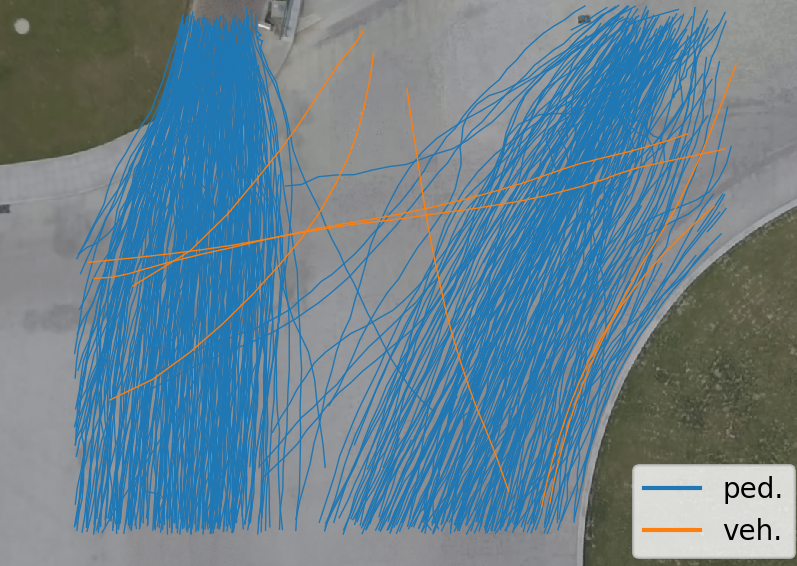}}
	\hfil
	\subfloat[Intersection]{\includegraphics[width=1.5in]{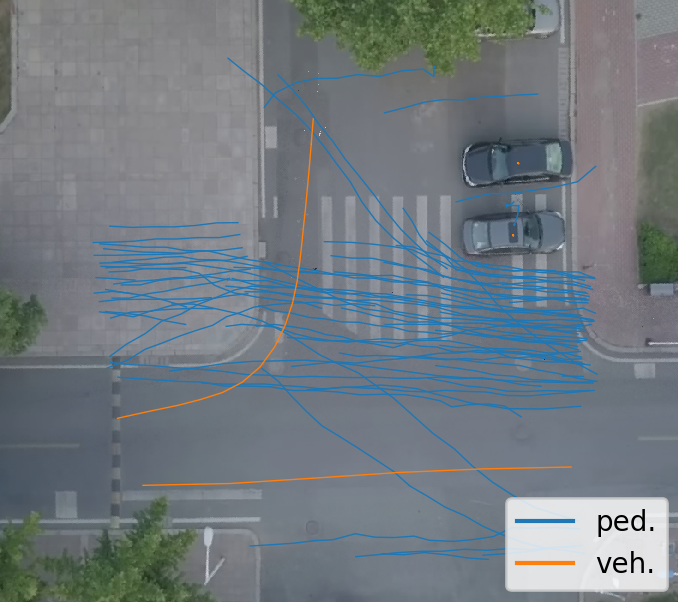}}
	\caption{Mixed trajectories from the DUT dataset.}
	\label{fig:dutdataset}
\end{figure}
To examine the transferability of our model, we validate the model by the scenarios extracted from a dataset from Dalian University of Technology campus in China \cite{yang2019top} (labelled DUT in the following). We observe that the DUT scenario differs from the HBS dataset in terms of environment layout, traffic conditions, and the social norms followed by the road users. The DUT dataset includes trajectories of 1793 pedestrians and 69 vehicles from a roundabout and an intersection, 11 and 17 video clips, respectively. We extracted 37 car-to-crowd interaction scenarios from the DUT dataset, which is less in number than the ones from HBS because of the short length of the clips from DUT. However, these scenarios contain a larger set of participants compared to the HBS scenarios.

\begin{figure}[htbp]
	\centering
	\vspace{-1.15em}
	\subfloat[Trajectory Difference ]{\includegraphics[width=2.83in]{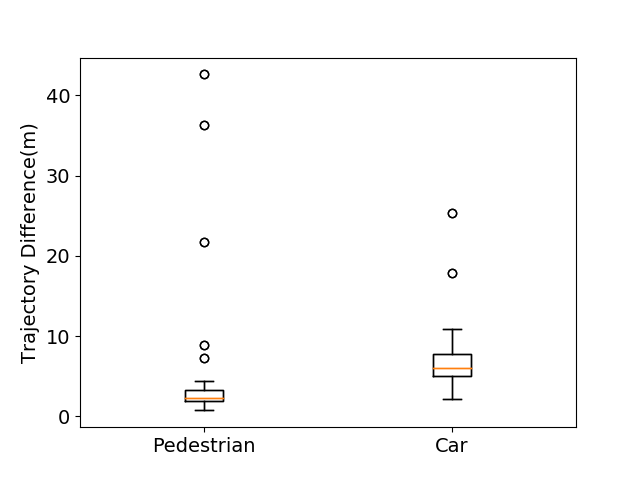}}
	\hfil
	\subfloat[Speed differences ]{\includegraphics[width=2.6in]{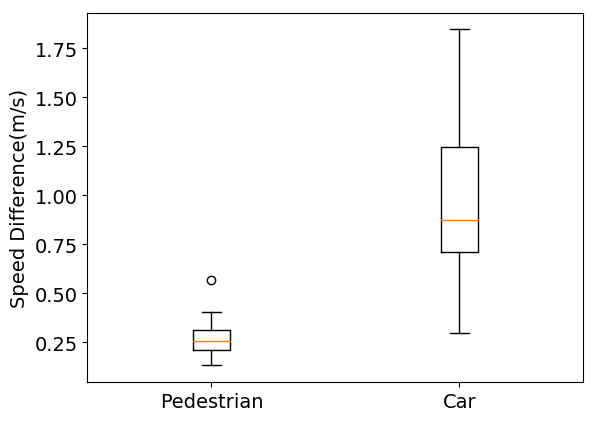}}
	\caption{The difference in trajectory and speed of real and simulated
pedestrians and cars of the DUT dataset.}
	\label{fig:QuantitativeDongfangbefore}
\end{figure}

Figure \ref{fig:QuantitativeDongfangbefore} shows the performance of our calibrated model (using the HBS dataset in Section \ref{sec:calibration-methodology}) on the DUT dataset. 
The mean and standard deviation in the difference of real and simulated
speeds of all road users involved in all the DUT scenarios are:  0.9835$ms^{-0.5}$ and 0.4371$ms^{-0.5}$ (cars); 0.2633 $ms^{-0.5}$ and 0.0947$ms^{-0.5}$ (pedestrians); and the mean and standard deviation in the difference of trajectories are: 6.8976 $m$ and 4.3551 $m$ (cars); 5.6098 $m$ and 9.4942 $m$ (pedestrians).

\begin{figure*}[!htbp]
	\centering
     \includegraphics[width=6in]{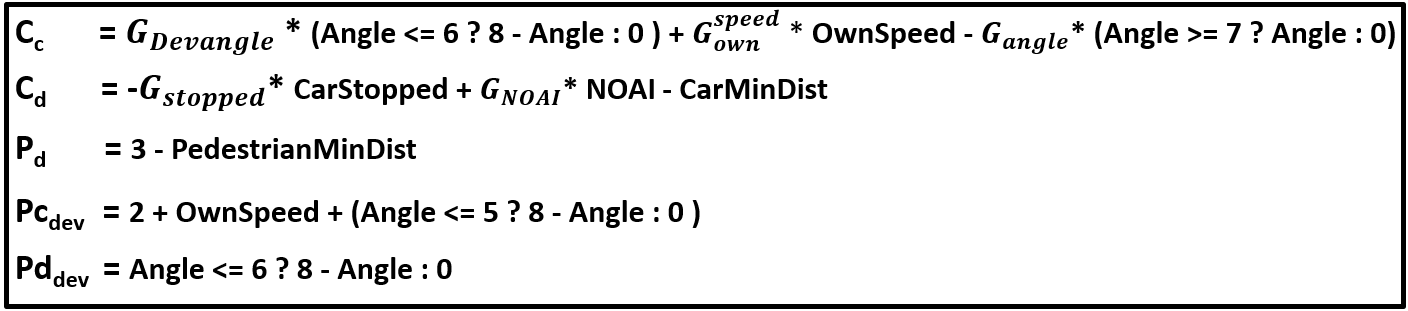}
    
	\caption{The impacts of situation dynamic on car-to-crowd interactions from the DUT Dataset.}
	\label{gamematrixdongfang}
\end{figure*}

Figure \ref{fig:QuantitativeDongfangbefore} shows that the performance of our model drops for the new data set, as the behaviour of road users in the HBS and DUT datasets differs, following are some findings in difference of their behaviours:
\begin{itemize}
    \item In the DUT dataset, road users maintain less inter-distance (i.e. Safety distance) compared to the HBS dataset.
    \item In the HBS dataset, the car drivers more often give pedestrians the priority to pass, whereas in the DUT dataset, pedestrians prefer to stop or deviate from their predefined path unless they are walking very first.
    \item The impact of the selected features in Subsection \ref{sec:calibration-methodology}.\ref{subsec:gamecalibration} is different in different datasets, as an example, CarStopped in the HBS dataset has a large influence in car's decision-making process. If CarStopped is 1, than the car most likely to stop. However, in the DUT dataset, CarStopped feature has the opposite influence in car's decision-making. 
\end{itemize}
We try to add these differences to our model, so that it can cope with the new situation more realistically. Figure \ref{gamematrixdongfang} shows the adjusted way to calculate payoff matrix for the DUT dataset.  For the DUT dataset, the \textit{MinDist} feature is also modified and divide into two parts: \textbf{PedestrianMinDist:} has value distance($\alpha, \beta$), if manhattanDistance($\alpha, \beta$) < $D_{min}$ and distance($\alpha, \beta$) - manhattanDistance($\alpha, \beta$) $\leq$ $M$, otherwise 0 and \textbf{CarMinDist:} has value manhattanDistance($\alpha, \beta$), if manhattanDistance($\alpha, \beta$) < $N$ and $S^\alpha_{current}$ > $S^\alpha_{high}$ ($\alpha$ as a pedestrian), otherwise 0. We denote the euclidean distance between $\alpha$ and $\beta$ as distance($\alpha, \beta$).

\begin{figure}[htbp]
	\centering
	\vspace{-1.15em}
	\subfloat[Trajectory Difference ]{\includegraphics[width=2.8in]{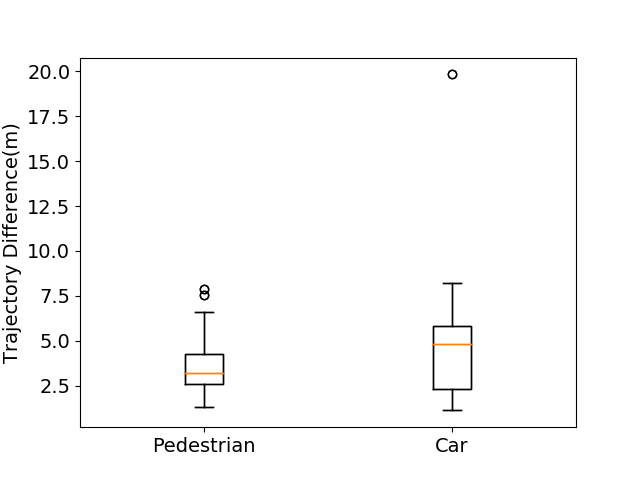}}
	\hfil
	\subfloat[Speed differences ]{\includegraphics[width=2.6in]{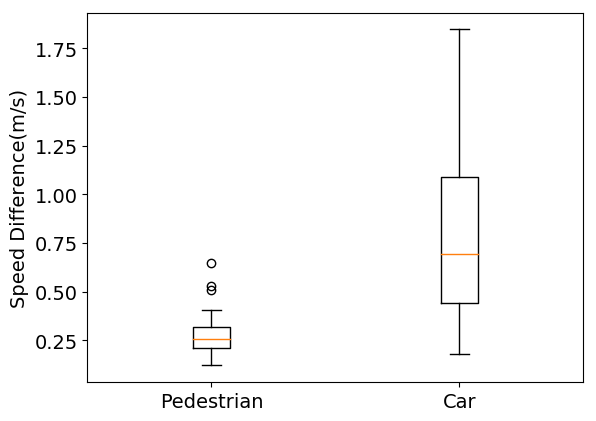}}
	\caption{The difference in trajectory and speed of real and simulated road users of the DUT dataset on the adjusted model.}
	\label{fig:QualitativeDongfang}
    \vspace{-0.75em}
\end{figure}
Figure \ref{fig:QualitativeDongfang} shows the performance of the adjusted model for the DUT dataset on the DUT scenarios. 
The mean and standard deviation in the difference of real and simulated
speeds are: 0.8263 $ms^{-0.5}$ and 0.472$ms^{-0.5}$ (cars); 0.2787$ms^{-0.5}$ and 0.1155$ms^{-0.5}$ (pedestrians) and the mean and standard deviation in the difference of trajectories are: 4.774 $m$ and 3.39 $m$ (cars); 
3.691 $m$ and 1.6837 $m$ (pedestrians). The results show a large improvement in the trajectory modelling of both pedestrians and cars. Moreover, the speed performance of cars has slight improvement but pedestrian is slightly degraded. The model shows satisfactory improvement in performance by only adding some social norms in payoff estimation process, reducing the value of the view ranged $V_R$ and safety distance $D_{min}$ ( by  performing a sensitivity analysis similar the one in \cite{chen2018social}), and skipping $I_{stopping}$. Calibrating the model parameters, including new parameters $M$ and $N$, for the DUT dataset is part of our future work.

\section{Conclusion and Future Work}
\label{sec:conclusion}

In this paper, we presented a methodological approach to calibrate and validate agent-based simulation models for shared spaces involving pedestrians and vehicles, created with our Game-Theoretic Social Force Model (GSFM) modelling approach. We proposed a generic calibration method that uses a genetic algorithm to automatically calibrate model parameters, and uses backward elimination techniques to select the most relevant features for a given model for complexity reduction and performance improvement. 

We also analysed the transferability of our simulation model, i.e., the ability to successfully apply a shared space model extracted for one specific shared space environment to a different target environment, with only little modification and adaptation. On the one hand, our results confirm the rather obvious conjecture that applying a given model that is calibrated to a specific type of shared space environment, to an arbitrary different environment will lead to relatively poor results. On the other hand, our cases study reveals that by adding some social norms extracted for our target environment,  our initial model can be adapted with reasonable effort to the new environment, achieving satisfactory performance.

Thus, our study provides evidence that re-using shared-space models may be feasible across a large range of environments, given appropriate methodological support for calibration. Even though our calibration method has only been tested for GSFM modelling approach, we believe that, due to the general nature of GSFM and the calibration procedure described in Section~\ref{sec:calibration-methodology}, and through our experience in combining GSFM with other model types (e.g. deep learning based ones \cite{Johora2020agent}),
our results have good potential for generalisation to other modelling paradigms. Proving this is a task for future work.  In addition, our future research will focus on calibrating the game parameters for a wider range of road users (e.g., including cyclists) and interactions (e.g., vehicle-to-vehicle), and calibrating the complete model using the DUT and other open source datasets of shared spaces. Most importantly, we shall work on recognition and modelling of different navigation patterns of individual user types and studying larger scenarios to explore the scalability of various interaction types and our simulation model.

\section*{Acknowledgements}
This research has been supported by the German Research Foundation (DFG) through the Research Training Group SocialCars (GRK 1931).
The authors thank the participants of the DFG research project MODIS (DFG project \#248905318) for providing the HBS dataset.

\bibliographystyle{ieeetr}  
\bibliography{mybibliography} 

\begin{thebibliography}{10}

\bibitem{emma2006shared}
E.~Clarke, ``Shared space-: the alternative approach to calming traffic,'' {\em
  Traffic Engineering \& Control}, vol.~47, no.~8, pp.~290--292, 2006.

\bibitem{hamilton2008shared}
B.~Hamilton-Baillie, ``Shared space: Reconciling people, places and traffic,''
  {\em Built environment}, vol.~34, no.~2, pp.~161--181, 2008.

\bibitem{gettman2003surrogate}
D.~Gettman and L.~Head, ``Surrogate safety measures from traffic simulation
  models,'' {\em Transportation Research Record: Journal of the Transportation
  Research Board}, no.~1840, pp.~104--115, 2003.

\bibitem{danaf2020pedestrian}
M.~Danaf, A.~Sabri, M.~Abou-Zeid, and I.~Kaysi, ``Pedestrian--vehicular
  interactions in a mixed street environment,'' {\em Transportation Letters},
  vol.~12, no.~2, pp.~87--99, 2020.

\bibitem{zheng2017driver}
Y.~Zheng, R.~T. Chase, L.~Elefteriadou, V.~Sisiopiku, and B.~Schroeder,
  ``Driver types and their behaviors within a high level of pedestrian activity
  environment,'' {\em Transportation letters}, vol.~9, no.~1, pp.~1--11, 2017.

\bibitem{helbing1995social}
D.~Helbing and P.~Molnar, ``Social force model for pedestrian dynamics,'' {\em
  Physical review E}, vol.~51, no.~5, p.~4282, 1995.

\bibitem{schonauer2017microscopic}
R.~Sch{\"o}nauer, {\em A Microscopic Traffic Flow Model for Shared Space}.
\newblock PhD thesis, Graz University of Technology, 2017.

\bibitem{anvari2015modelling}
B.~Anvari, M.~G. Bell, A.~Sivakumar, and W.~Y. Ochieng, ``Modelling shared
  space users via rule-based social force model,'' {\em Transportation Research
  Part C: Emerging Technologies}, vol.~51, pp.~83--103, 2015.

\bibitem{rinke2017multi}
N.~Rinke, C.~Schiermeyer, F.~Pascucci, V.~Berkhahn, and B.~Friedrich, ``A
  multi-layer social force approach to model interactions in shared spaces
  using collision prediction,'' {\em Transportation Research Procedia},
  vol.~25, pp.~1249--1267, 2017.

\bibitem{lan2005inhomogeneous}
L.~W. Lan and C.-W. Chang, ``Inhomogeneous cellular automata modeling for mixed
  traffic with cars and motorcycles,'' {\em Journal of Advanced
  Transportation}, vol.~39, no.~3, pp.~323--349, 2005.

\bibitem{zhang2007modeling}
Y.~Zhang and H.~Duan, ``Modeling mixed traffic flow at crosswalks in
  micro-simulations using cellular automata,'' {\em Tsinghua Science \&
  Technology}, vol.~12, no.~2, pp.~214--222, 2007.

\bibitem{bandini2017collision}
S.~Bandini, L.~Crociani, C.~Feliciani, A.~Gorrini, and G.~Vizzari, ``Collision
  avoidance dynamics among heterogeneous agents: The case of pedestrian/vehicle
  interactions,'' in {\em Conference of the Italian Association for Artificial
  Intelligence}, pp.~44--57, Springer, 2017.

\bibitem{cheng2019pedestrian}
H.~Cheng, Y.~Li, and M.~Sester, ``Pedestrian group detection in shared space,''
  in {\em 2019 IEEE Intelligent Vehicles Symposium (IV)}, pp.~1707--1714, IEEE,
  2019.

\bibitem{johora2018modeling}
F.~T. Johora and J.~P. M{\"u}ller, ``Modeling interactions of multimodal road
  users in shared spaces,'' in {\em 2018 21st International Conference on
  Intelligent Transportation Systems (ITSC)}, pp.~3568--3574, IEEE, 2018.

\bibitem{bandini2017approach}
S.~Bandini, L.~Crociani, and G.~Vizzari, ``An approach for managing
  heterogeneous speed profiles in cellular automata pedestrian models.,'' {\em
  Journal of Cellular Automata}, vol.~12, no.~5, 2017.

\bibitem{burstedde2001simulation}
C.~Burstedde, K.~Klauck, A.~Schadschneider, and J.~Zittartz, ``Simulation of
  pedestrian dynamics using a two-dimensional cellular automaton,'' {\em
  Physica A: Statistical Mechanics and its Applications}, vol.~295, no.~3-4,
  pp.~507--525, 2001.

\bibitem{nagel1992cellular}
K.~Nagel and M.~Schreckenberg, ``A cellular automaton model for freeway
  traffic,'' {\em Journal de physique I}, vol.~2, no.~12, pp.~2221--2229, 1992.

\bibitem{chai2015fuzzy}
C.~Chai and Y.~D. Wong, ``Fuzzy cellular automata model for signalized
  intersections,'' {\em Computer-Aided Civil and Infrastructure Engineering},
  vol.~30, no.~12, pp.~951--964, 2015.

\bibitem{chen2018evaluating}
J.~Chen, Z.~Li, W.~Wang, and H.~Jiang, ``Evaluating bicycle--vehicle conflicts
  and delays on urban streets with bike lane and on-street parking,'' {\em
  Transportation letters}, vol.~10, no.~1, pp.~1--11, 2018.

\bibitem{chen2018social}
X.~Chen, M.~Treiber, V.~Kanagaraj, and H.~Li, ``Social force models for
  pedestrian traffic--state of the art,'' {\em Transport reviews}, vol.~38,
  no.~5, pp.~625--653, 2018.

\bibitem{asano2010microscopic}
M.~Asano, T.~Iryo, and M.~Kuwahara, ``Microscopic pedestrian simulation model
  combined with a tactical model for route choice behaviour,'' {\em
  Transportation Research Part C: Emerging Technologies}, vol.~18, no.~6,
  pp.~842--855, 2010.

\bibitem{johora2017dynamic}
F.~T. Johora, P.~Kraus, and J.~P. M{\"u}ller, ``Dynamic path planning and
  movement control in pedestrian simulation,'' {\em arXiv preprint
  arXiv:1709.08235}, 2017.

\bibitem{yang2018social}
D.~Yang, {\"U}.~{\"O}zg{\"u}ner, and K.~Redmill, ``Social force based
  microscopic modeling of vehicle-crowd interaction,'' in {\em 2018 IEEE
  Intelligent Vehicles Symposium (IV)}, pp.~1537--1542, IEEE, 2018.

\bibitem{zeng2014modified}
W.~Zeng, H.~Nakamura, and P.~Chen, ``A modified social force model for
  pedestrian behavior simulation at signalized crosswalks,'' {\em
  Procedia-Social and Behavioral Sciences}, vol.~138, pp.~521--530, 2014.

\bibitem{pascucci2018should}
F.~Pascucci, N.~Rinke, C.~Schiermeyer, V.~Berkhahn, and B.~Friedrich, ``Should
  i stay or should i go? a discrete choice model for pedestrian--vehicle
  conflicts in shared space,'' tech. rep., 2018.

\bibitem{fujii2017agent}
H.~Fujii, H.~Uchida, and S.~Yoshimura, ``Agent-based simulation framework for
  mixed traffic of cars, pedestrians and trams,'' {\em Transportation research
  part C: emerging technologies}, vol.~85, pp.~234--248, 2017.

\bibitem{lutteken2016using}
N.~L{\"u}tteken, M.~Zimmermann, and K.~J. Bengler, ``Using gamification to
  motivate human cooperation in a lane-change scenario,'' in {\em Intelligent
  Transportation Systems (ITSC), 2016 IEEE 19th International Conference on},
  pp.~899--906, IEEE, 2016.

\bibitem{bjornskau2017zebra}
T.~Bj{\o}rnskau, ``The zebra crossing game--using game theory to explain a
  discrepancy between road user behaviour and traffic rules,'' {\em Safety
  science}, vol.~92, pp.~298--301, 2017.

\bibitem{aschermann2016lightjason}
M.~Aschermann, P.~Kraus, and J.~P. M{\"u}ller, ``{LightJason: A BDI Framework
  inspired by Jason},'' in {\em Multi-Agent Systems and Agreement Technologies:
  14th Europ. Conf., EUMAS 2016}, vol.~10207 of {\em {LNCS}}, pp.~58--66,
  Springer, 2017.

\bibitem{koefoed2012representations}
A.~Koefoed-Hansen and G.~S. Brodal, {\em Representations for Path Finding in
  Planar Environments}.
\newblock PhD thesis, Citeseer, 2012.

\bibitem{millington2009artificial}
I.~Millington and J.~Funge, {\em Artificial intelligence for games}.
\newblock CRC Press, 2009.

\bibitem{johansson2008specification}
A.~Johansson, D.~Helbing, and P.~K. Shukla, ``Specification of a microscopic
  pedestrian model by evolutionary adjustment to video tracking data [c],''
  {\em Advances in Complex System{\copyright} World Scientific Publishing
  Company}, vol.~25, 2008.

\bibitem{pascucci2017discrete}
F.~Pascucci, N.~Rinke, C.~Schiermeyer, V.~Berkhahn, and B.~Friedrich, ``A
  discrete choice model for solving conflict situations between pedestrians and
  vehicles in shared space,'' {\em arXiv preprint arXiv:1709.09412}, 2017.

\bibitem{zames1981genetic}
G.~Zames, N.~Ajlouni, N.~Ajlouni, N.~Ajlouni, J.~Holland, W.~Hills, and
  D.~Goldberg, ``Genetic algorithms in search, optimization and machine
  learning.,'' {\em Information Technology Journal}, vol.~3, no.~1,
  pp.~301--302, 1981.

\bibitem{amirjamshidi2019multi}
G.~Amirjamshidi and M.~J. Roorda, ``Multi-objective calibration of traffic
  microsimulation models,'' {\em Transportation Letters}, vol.~11, no.~6,
  pp.~311--319, 2019.

\bibitem{cunha2009genetic}
A.~L. Cunha, J.~E. Bessa, and J.~R. Setti, ``Genetic algorithm for the
  calibration of vehicle performance models of microscopic traffic
  simulators,'' in {\em Portuguese Conference on Artificial Intelligence},
  pp.~3--14, Springer, 2009.

\bibitem{garcia2014process}
B.~Garc{\'\i}a~de Soto, B.~T. Adey, and D.~Fernando, ``A process for the
  development and evaluation of preliminary construction material quantity
  estimation models using backward elimination regression and neural
  networks,'' {\em Journal of Cost Analysis and Parametrics}, vol.~7, no.~3,
  pp.~180--218, 2014.

\bibitem{yang2019top}
D.~Yang, L.~Li, K.~Redmill, and {\"U}.~{\"O}zg{\"u}ner, ``Top-view
  trajectories: A pedestrian dataset of vehicle-crowd interaction from
  controlled experiments and crowded campus,'' {\em arXiv preprint
  arXiv:1902.00487}, 2019.

\bibitem{Johora2020agent}
F.~T. Johora, H.~Cheng, J.~P. M{\"u}ller, and M.~Sester, ``An extended abstract
  on: An agent-based model for trajectory modelling in shared spaces: A
  combination of expert-based and deep learning approaches,'' in {\em AAMAS},
  ACM, 2020.
\newblock (accepted).

\end{thebibliography}

\listoffigures
\listoftables
\end{document}